\documentclass[conference]{IEEEtran}

\IEEEoverridecommandlockouts
\IEEEpubid{\makebox[\columnwidth]{ }\hspace{\columnsep}\makebox[\columnwidth]{ }}

\usepackage[T1]{fontenc}
\usepackage{cite}
\usepackage{amsmath,amssymb,amsfonts}
\usepackage{bm}
\usepackage{booktabs}
\usepackage[pdftex]{graphicx}
\usepackage{algorithm}
\usepackage{algorithmic}
\usepackage{balance}

\graphicspath{{figures/}{./}}

\newcommand{\mat}[1]{\mathbf{#1}}
\newcommand{\vect}[1]{\boldsymbol{#1}}
\newcommand{\E}{\mathbb{E}}
\newcommand{\R}{\mathbb{R}}
\DeclareMathOperator{\tr}{tr}

\DeclareMathOperator{\vecop}{vec}

\newcommand{\safeincludegraphics}[2][]{%
  \IfFileExists{#2}{%
    \includegraphics[#1]{#2}%
  }{%
    \fbox{\begin{minipage}[c][0.20\textheight][c]{0.90\linewidth}
    \centering Figure file not found:\\[2pt]\texttt{\detokenize{#2}}
    \end{minipage}}%
  }%
}

\begin{document}

\title{Pilot-Aided MIMO Channel Identification and Linear Deconvolution in Correlated Gaussian Noise}

\author{\IEEEauthorblockN{Necati Kagan Erkek \IEEEauthorrefmark{2}, Y. Ugur Ozcan\IEEEauthorrefmark{2}}
\IEEEauthorblockA{\IEEEauthorrefmark{2} Telecommunications Engineering, Department of Electronic, Information and Bioengineering \\ Politecnico di Milano, Piazza Leonardo da Vinci 32, 20133, Milan, Italy} \texttt{necatikagan.erkek@mail.polimi.it, yilmazugur.ozcan@mail.polimi.it}}

\maketitle

\begin{abstract}
This paper presents a pilot-aided study of multiple-input multiple-output (MIMO) channel identification and linear deconvolution under spatially correlated Gaussian noise. A real-valued $4\times4$ baseband model is analyzed for both memoryless and finite-impulse-response channels. The noise process is generated from a Toeplitz covariance matrix, the channel is estimated from pilot symbols through maximum-likelihood/least-squares formulations, and the empirical mean-square error is compared with the Cramer--Rao bound. The estimated channel is then used for data-symbol recovery through maximum-likelihood zero-forcing and linear minimum-mean-square-error deconvolution. The results show that sufficiently long and well-conditioned pilot blocks allow the channel estimator to approach the theoretical lower bound, whereas short training intervals cause rank and conditioning limitations, especially for the four-tap model. The deconvolution experiments further show that MMSE regularization provides a more stable inverse than unregularized zero forcing at low signal-to-noise ratios and for inaccurate channel estimates.
\end{abstract}
\vspace{8pt}
\vspace{4pt}
\section{Introduction}

Multiple-input multiple-output (MIMO) processing is a fundamental technique for modern wireless communications and multichannel signal processing because it exploits several transmit and receive dimensions to increase spatial multiplexing capability, diversity gain, and channel observability. The theoretical and practical foundations of MIMO communication were established through space-time coding, layered architectures, and capacity analyses for multi-antenna Gaussian channels \cite{alamouti,tarokh,foschini,telatar,goldsmith,paulraj,biglieri,tse}. Later developments in large-scale and massive MIMO systems further emphasized the importance of reliable channel-state information when multiple spatial streams are separated at the receiver \cite{rusek,larsson,bjornson}. In all of these systems, the quality of the estimated channel matrix has a direct effect on equalization, interference suppression, and data-symbol recovery.

Pilot-aided channel estimation is therefore a central component of practical MIMO receivers. Training sequences must provide enough independent observations to identify the unknown channel coefficients while preserving a reasonable fraction of the frame for payload data. The structure and length of the pilot block have been studied extensively for multiple-antenna systems, including BLAST-type training and optimal training-length analyses \cite{marzetta,hassibi}. The statistical tools used in this context are based on classical estimation theory, where maximum-likelihood (ML), least-squares (LS), generalized least-squares, and Bayesian linear estimators provide systematic methods for recovering unknown parameters from noisy observations \cite{cramer,rao,kay,van_trees,poor,kailath,ljung,soderstrom,scharf}. The Cramer--Rao bound (CRB) is especially useful because it supplies an analytical lower bound on the variance of unbiased estimators and makes it possible to interpret whether a simulation result is limited by the estimator itself or by the information content of the pilot block. The covariance and vectorized-matrix formulations used in this paper also follow standard Toeplitz, Kronecker-product, and matrix-analysis arguments \cite{gray,brewer,horn}.

The data-recovery stage is an inverse problem that must also be treated carefully. A zero-forcing or ML inverse can remove the deterministic channel effect when the estimated channel is full rank, but it may amplify noise and channel-estimation errors when the channel matrix is ill conditioned. Linear minimum-mean-square-error (MMSE) deconvolution introduces a regularization term that balances inversion accuracy against noise enhancement, a principle that is central in digital communications, adaptive filtering, Wiener filtering, and statistical signal processing \cite{proakis,verdu,wiener,haykin,sayed,oppenheim,qureshi,widrow}. Numerical conditioning is also essential because all LS and inverse-filtering procedures rely on matrix factorizations or normal equations whose stability depends on the singular-value structure of the pilot and channel matrices \cite{tikhonov,hansen,golub,bjorck}.

This paper reformulates a MATLAB-based study into a formal conference-paper presentation of MIMO system identification and deconvolution. The study pursues three objectives: validating a correlated Gaussian noise generator through sample covariance experiments, quantifying the accuracy of pilot-aided channel estimation for memoryless and four-tap $4\times4$ channels, and evaluating how the estimated channel affects ML and MMSE data-symbol reconstruction. The main contribution is a compact but complete simulation framework that links covariance validation, CRB-based channel-identification assessment, and pilot-length selection for linear deconvolution. The paper also clarifies the role of pilot rank, channel memory, noise correlation, and inverse regularization in the final reconstruction error.
\\

The rest of the paper is organized as follows. Section~\ref{sec:model} introduces the MIMO signal model and the correlated-noise covariance structure. Section~\ref{sec:methodology} derives the channel-estimation and deconvolution formulations. Section~\ref{sec:simulation} summarizes the simulation parameters. Section~\ref{sec:results} presents and discusses the numerical results. Section~\ref{sec:conclusion} concludes the paper.

\section{System Model}
\label{sec:model}

Consider a real-valued discrete-time MIMO channel with $N$ transmit streams and $M$ receive streams. The received signal at output $i$ is modeled as
\begin{equation}
    y_i[k] = \sum_{j=1}^{N}\sum_{\ell=0}^{L_h-1} h_{ij}[\ell]x_j[k-\ell] + w_i[k],
    \label{eq:scalar_model}
\end{equation}
where $x_j[k]$ is the transmitted symbol on input $j$, $h_{ij}[\ell]$ is the channel impulse response from input $j$ to output $i$, $L_h$ is the channel memory length, and $w_i[k]$ is additive noise. In vector form, \eqref{eq:scalar_model} becomes
\begin{equation}
    \mat y[k] = \sum_{\ell=0}^{L_h-1} \mat H[\ell]\mat x[k-\ell] + \mat w[k],
    \label{eq:vector_model}
\end{equation}
where $\mat y[k]\in\R^M$, $\mat x[k]\in\R^N$, and $\mat H[\ell]\in\R^{M\times N}$. The numerical study uses $M=N=4$, which gives a square spatial channel with sixteen coefficients in the memoryless case and sixty-four coefficients in the four-tap case.

The spatial noise vector is assumed to be zero-mean Gaussian with covariance
\begin{equation}
    \mat w[k]\sim \mathcal{N}(\mat 0,\mat C_w), \qquad
    [\mat C_w]_{ij}=\sigma_w^2\rho^{|i-j|},
    \label{eq:noise_model}
\end{equation}
where $\rho$ is the inter-channel correlation coefficient and $|\rho|\leq 1$. The covariance in \eqref{eq:noise_model} is Toeplitz, so adjacent receive branches have stronger correlation than distant branches. In the simulations, correlated noise is generated by
\begin{equation}
    \mat w[k]=\mat C_w^{1/2}\mat n[k], \qquad \mat n[k]\sim\mathcal{N}(\mat 0,\mat I_M),
    \label{eq:noise_generation}
\end{equation}
where $\mat C_w^{1/2}$ is obtained numerically by a matrix square root. This construction gives $\E\{\mat w[k]\mat w^T[k]\}=\mat C_w$.

Two channel models are evaluated. The first is the memoryless model, $L_h=1$, for which the channel is represented by a single matrix $\mat H$. The second is the four-tap finite-impulse-response (FIR) model, $L_h=4$, with deterministic coefficients
\begin{equation}
    h_{ij}[\ell] = \alpha^{|i-j|}\beta^{\ell}, \qquad \ell=0,1,2,3.
    \label{eq:fir_channel}
\end{equation}
The parameter $\alpha$ controls spatial coupling between different transmit and receive branches, whereas $\beta$ controls the temporal decay across taps. This model is intentionally simple, but it is useful for separating the effects of spatial coupling, memory length, and pilot excitation on the estimator variance.

\section{Estimation and Deconvolution Methodology}
\label{sec:methodology}

\subsection{Memoryless Channel Identification}
\label{subsec:memoryless_estimator}

Let $\mat X_Q=[\mat x[1],\ldots,\mat x[Q]]\in\R^{N\times Q}$ denote the pilot matrix and $\mat Y_Q=[\mat y[1],\ldots,\mat y[Q]]\in\R^{M\times Q}$ the received pilot block. For $L_h=1$, the block input-output relation is
\begin{equation}
    \mat Y_Q = \mat H\mat X_Q + \mat W_Q.
    \label{eq:memoryless_block}
\end{equation}
When $\mat X_Q\mat X_Q^T$ is nonsingular and the noise is Gaussian, the ML estimate is identical to the LS estimate:
\begin{equation}
    \widehat{\mat H}_{\mathrm{ML}} = \mat Y_Q\mat X_Q^T(\mat X_Q\mat X_Q^T)^{-1}.
    \label{eq:ml_H}
\end{equation}
The estimator in \eqref{eq:ml_H} is unbiased for a deterministic channel under the assumed noise model. The vectorized observation model is
\begin{equation}
    \vecop(\mat Y_Q) = (\mat X_Q^T\otimes \mat I_M)\vecop(\mat H) + \vecop(\mat W_Q),
    \label{eq:vectorized_memoryless}
\end{equation}
with $\operatorname{Cov}(\vecop(\mat W_Q))=\mat I_Q\otimes \mat C_w$. Therefore, the Fisher information matrix for $\vecop(\mat H)$ is
\begin{equation}
    \mat J_H = (\mat X_Q\mat X_Q^T)\otimes \mat C_w^{-1},
    \label{eq:fim_memoryless}
\end{equation}
which gives the CRB
\begin{equation}
    \operatorname{Cov}\!\left(\vecop(\widehat{\mat H})\right)
    \succeq (\mat X_Q\mat X_Q^T)^{-1}\otimes \mat C_w.
    \label{eq:crb_memoryless}
\end{equation}
The corresponding normalized lower bound on the channel-estimation MSE is
\begin{equation}
    \mathrm{CRB}_{H}=\frac{1}{MN}\tr\!\left((\mat X_Q\mat X_Q^T)^{-1}\right)\tr(\mat C_w).
    \label{eq:crb_mse_memoryless}
\end{equation}
This expression explicitly shows that a longer and better-conditioned pilot block reduces the achievable estimation error.

\subsection{Four-Tap Channel Identification}
\label{subsec:fir_estimator}

For $L_h>1$, the pilot sequence must be expanded into a convolution matrix that contains delayed versions of all transmit streams. Let $\vect\theta$ collect all $MNL_h$ coefficients in $\{\mat H[0],\ldots,\mat H[L_h-1]\}$, and let $\mat\Phi_Q$ denote the block convolution matrix constructed from the pilot samples. The valid received pilot samples can then be stacked as
\begin{equation}
    \mat y_Q = \mat \Phi_Q\vect\theta + \mat w_Q.
    \label{eq:fir_block}
\end{equation}
If $\mat\Sigma_w$ is the covariance matrix of the stacked noise vector, the Gaussian ML estimator is the generalized LS solution
\begin{equation}
    \widehat{\vect\theta}_{\mathrm{ML}} =
    (\mat \Phi_Q^T\mat \Sigma_w^{-1}\mat \Phi_Q)^{-1}
    \mat \Phi_Q^T\mat \Sigma_w^{-1}\mat y_Q.
    \label{eq:fir_gls}
\end{equation}
The associated CRB is
\begin{equation}
    \operatorname{Cov}(\widehat{\vect\theta}) \succeq
    (\mat \Phi_Q^T\mat \Sigma_w^{-1}\mat \Phi_Q)^{-1}.
    \label{eq:fir_crb}
\end{equation}
The rank of $\mat\Phi_Q$ is a decisive factor. Each receive branch contains $NL_h$ unknown tap coefficients; consequently, the four-input, four-tap case requires at least sixteen independent equations per receive branch. Pilot blocks with $Q$ close to or below this threshold can lead to singular or poorly conditioned normal matrices. This explains why FIR channel estimation is more sensitive to pilot length than the memoryless case.

\subsection{Linear Deconvolution}
\label{subsec:deconvolution_method}

After channel identification, the receiver uses the estimated memoryless channel to recover data symbols. For a data sample,
\begin{equation}
    \mat y[k] = \widehat{\mat H}\mat x[k] + \mat w[k].
    \label{eq:data_model}
\end{equation}
The ML, or weighted zero-forcing, estimate is
\begin{equation}
    \widehat{\mat x}_{\mathrm{ML}}[k]
    = (\widehat{\mat H}^{T}\mat C_w^{-1}\widehat{\mat H})^{-1}
    \widehat{\mat H}^{T}\mat C_w^{-1}\mat y[k].
    \label{eq:zf_deconv}
\end{equation}
This expression exactly inverts the estimated channel in a weighted LS sense, but it can amplify noise when $\widehat{\mat H}$ has small singular values. The linear MMSE estimate introduces a source covariance prior $\mat R_x$:
\begin{equation}
    \widehat{\mat x}_{\mathrm{MMSE}}[k]
    = (\widehat{\mat H}^{T}\mat C_w^{-1}\widehat{\mat H}+\mat R_x^{-1})^{-1}
    \widehat{\mat H}^{T}\mat C_w^{-1}\mat y[k].
    \label{eq:mmse_deconv}
\end{equation}
The simulations use independent unit-variance source symbols, so $\mat R_x=\mat I_N$. The added identity matrix in \eqref{eq:mmse_deconv} regularizes the inverse and limits noise enhancement.

The performance metrics are
\begin{align}
    \mathrm{MSE}_H &= \frac{1}{MNL_h}\sum_{\ell=0}^{L_h-1}
    \left\|\widehat{\mat H}[\ell]-\mat H[\ell]\right\|_F^2,\label{eq:mse_H}\\
    \mathrm{MSE}_X &= \frac{1}{N(P-Q)}\left\|\widehat{\mat X}_d-\mat X_d\right\|_F^2,
    \label{eq:mse_X}
\end{align}
where $P$ is the total frame length and $P-Q$ is the number of data samples per stream. A scalar pilot-selection metric is also used:
\begin{equation}
    J(Q)=\mathrm{MSE}_H+\mathrm{MSE}_X.
    \label{eq:decision_metric}
\end{equation}
This metric favors pilot lengths that reduce both channel-estimation error and data-reconstruction error. It does not include the throughput loss from pilot overhead, so it should be interpreted as an accuracy-oriented design indicator.

\section{Simulation Setup}
\label{sec:simulation}

The numerical experiments were implemented in MATLAB and averaged over independent Monte Carlo trials. The same general simulation procedure is used throughout the paper: a channel matrix or channel impulse response is generated from the selected $(\alpha,\beta)$ parameters, pilot and data symbols are drawn from independent standard Gaussian distributions, correlated noise is generated according to \eqref{eq:noise_generation}, and the empirical errors are averaged over repeated trials. Table~\ref{tab:parameters} summarizes the main parameters.

\begin{table}[!t]
\centering
\caption{Simulation parameters.}
\label{tab:parameters}
\footnotesize
\begin{tabular}{ll}
\toprule
Parameter & Value \\
\midrule
Transmit/receive dimensions & $N=M=4$ \\
Channel memory & $L_h=1$ and $L_h=4$ \\
SNR range & $-10$ to $30$ dB \\
Pilot lengths for channel estimation & $Q\in\{10,20,30,40,50\}$ \\
Pilot lengths for deconvolution & $Q=10,20,\ldots,200$ \\
Total frame length for deconvolution & $P=200$ samples per stream \\
Spatial coupling & $0\leq \alpha\leq 0.99$ \\
Noise correlation & $0\leq \rho\leq 0.99$ \\
Memory decay values & $\beta\in\{0.1,0.5,0.9\}$ \\
Monte Carlo trials & 500 for covariance validation \\
\bottomrule
\end{tabular}
\end{table}

The signal-to-noise ratio is controlled by scaling the noise covariance as $\mat C_w/\gamma$, where $\gamma=10^{\mathrm{SNR}/10}$. The empirical MSE curves are plotted against SNR so that the slope and separation of the curves can be compared directly with the CRB. For the deconvolution study, the total frame length is fixed at $P=200$. Increasing $Q$ therefore improves the channel estimate but leaves fewer samples for data reconstruction, which creates a practical training-overhead tradeoff.

\section{Numerical Results and Discussion}
\label{sec:results}

\subsection{Validation of the Correlated-Noise Generator}
\label{subsec:noise_generation}

Algorithm~\ref{alg:noise_validation} describes the procedure used to generate spatially correlated Gaussian noise and to validate its covariance. The covariance matrix is first built from the Toeplitz correlation law in \eqref{eq:noise_model}. White Gaussian samples are then transformed by $\mat C_w^{1/2}$, and the sample covariance is computed from the generated record. The reported MSE is the element-wise squared difference between the sample covariance and the prescribed covariance, averaged across Monte Carlo repetitions. This validation step is important because all later estimation and deconvolution experiments rely on the assumed noise covariance.

\begin{algorithm}[!hb]
\caption{Correlated-noise generation and covariance validation}
\label{alg:noise_validation}
\footnotesize
\begin{algorithmic}[1]
\STATE \textbf{Input:} antenna dimension $M$, correlation grid $\mathcal{R}$, sample-length grid $\mathcal{L}$, number of trials $N_{\mathrm{MC}}$
\STATE \textbf{Output:} covariance mismatch $\mathrm{MSE}_{C}(\rho,L)$
\FOR{each $\rho\in\mathcal{R}$}
    \STATE Form $[\mat C_w]_{ij}=\rho^{|i-j|}$ for $i,j=1,\ldots,M$
    \STATE Compute a square-root factor $\mat A=\mat C_w^{1/2}$
    \FOR{each $L\in\mathcal{L}$}
        \FOR{$m=1$ to $N_{\mathrm{MC}}$}
            \STATE Draw $\mat N_m\sim\mathcal{N}(\mat 0,\mat I)$ with size $M\times L$
            \STATE Generate correlated noise $\mat W_m=\mat A\mat N_m$
            \STATE Estimate $\widehat{\mat C}_{w,m}=\mat W_m\mat W_m^T/L$
            \STATE Compute $e_m=\|\widehat{\mat C}_{w,m}-\mat C_w\|_F^2/M^2$
        \ENDFOR
        \STATE Set $\mathrm{MSE}_{C}(\rho,L)=N_{\mathrm{MC}}^{-1}\sum_m e_m$
    \ENDFOR
\ENDFOR
\STATE \textbf{return} $\mathrm{MSE}_{C}(\rho,L)$
\end{algorithmic}
\end{algorithm}

\begin{figure}[!t]
\centering
\safeincludegraphics[width=0.95\linewidth]{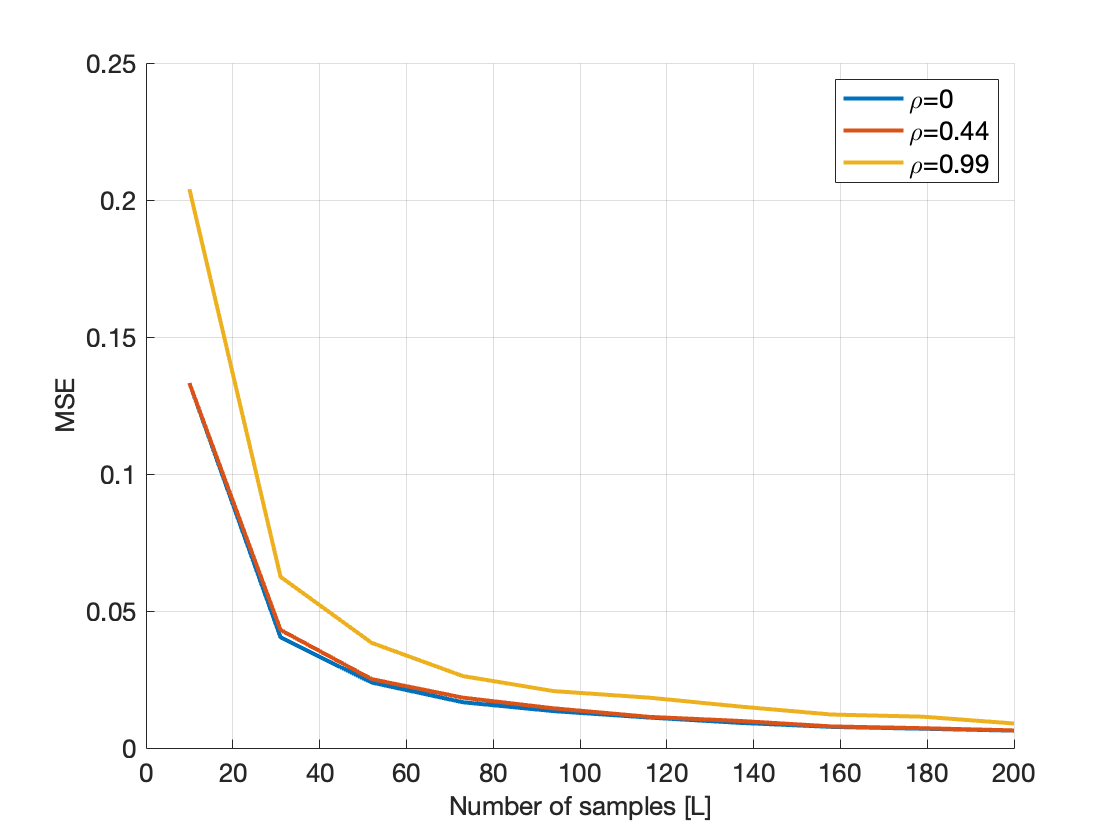}
\caption{Noise-covariance validation versus the number of generated samples. The curves report the MSE between the prescribed covariance matrix and the sample covariance matrix for several values of $\rho$. The figure demonstrates that increasing the record length produces a more accurate covariance estimate, while stronger noise correlation leads to a larger finite-sample mismatch for the same record length.}
\label{fig:noise_samples}
\end{figure}

Fig.~\ref{fig:noise_samples} shows the convergence of the sample covariance estimate as the number of generated samples increases. The MSE decreases rapidly when the record length grows from a very short block to a moderate block, which is consistent with the averaging effect of independent samples. The curves also indicate that high correlation is more difficult to estimate accurately from a finite record because the off-diagonal covariance entries become larger and more numerous. This result confirms that the noise-generation step is statistically consistent with the target covariance, while also showing that covariance validation should not be performed with very short sample sequences.

\begin{figure}[!t]
\centering
\safeincludegraphics[width=0.95\linewidth]{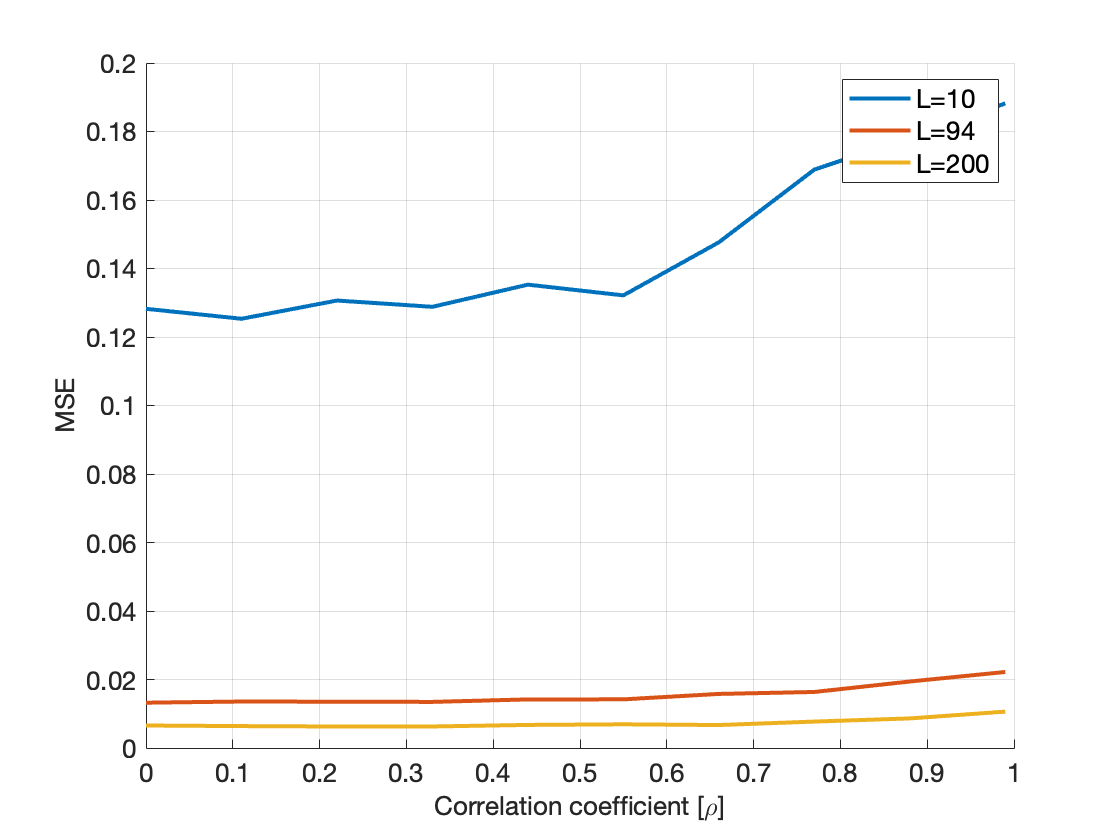}
\caption{Noise-covariance validation versus the correlation coefficient $\rho$ for different sample lengths. Short sample records exhibit a large covariance-estimation error over the full range of $\rho$, whereas longer records reduce the MSE and make the validation less sensitive to the off-diagonal covariance structure.}
\label{fig:noise_rho}
\end{figure}

Fig.~\ref{fig:noise_rho} presents the complementary view in which the correlation coefficient is swept for fixed sample lengths. For small $L$, the sample covariance contains substantial random fluctuations, so the MSE remains comparatively high. When $L$ increases, the covariance estimate becomes much closer to the theoretical matrix for all tested values of $\rho$. The figure also shows that the effect of $\rho$ becomes more visible when the available record is short. In the following channel-estimation experiments, this observation motivates the use of repeated Monte Carlo trials so that random covariance fluctuations do not dominate the estimator comparison.

\subsection{Memoryless Channel-Estimation Results}
\label{subsec:memoryless_results}

Algorithm~\ref{alg:memoryless_estimation} summarizes the memoryless channel-identification experiment. For each SNR and pilot length, a pilot matrix is generated, the received training block is formed, the ML/LS estimate is computed from \eqref{eq:ml_H}, and the empirical MSE is compared with the CRB from \eqref{eq:crb_mse_memoryless}. The algorithm also includes the channel-coupling parameter $\alpha$, which changes the deterministic entries of $\mat H$ but not the Fisher information matrix when the pilot matrix and noise covariance are fixed.

\begin{algorithm}[!t]
\caption{Memoryless MIMO channel estimation and CRB evaluation}
\label{alg:memoryless_estimation}
\footnotesize
\begin{algorithmic}[1]
\STATE \textbf{Input:} SNR grid $\Gamma$, pilot grid $\mathcal{Q}$, coupling values $\mathcal{A}$, covariance $\mat C_w$
\STATE \textbf{Output:} empirical $\mathrm{MSE}_{H}$ and $\mathrm{CRB}_{H}$
\FOR{each $\gamma\in\Gamma$}
    \FOR{each $Q\in\mathcal{Q}$}
        \STATE Draw the pilot matrix $\mat X_Q\in\R^{N\times Q}$
        \FOR{each $\alpha\in\mathcal{A}$}
            \STATE Form the memoryless channel $[\mat H]_{ij}=\alpha^{|i-j|}$
            \STATE Draw $\mat W_Q\sim\mathcal{N}(\mat 0,\mat I_Q\otimes \mat C_w/\gamma)$
            \STATE Generate $\mat Y_Q=\mat H\mat X_Q+\mat W_Q$
            \STATE Estimate $\widehat{\mat H}=\mat Y_Q\mat X_Q^T(\mat X_Q\mat X_Q^T)^{-1}$
            \STATE Compute $\mathrm{MSE}_{H}=\|\widehat{\mat H}-\mat H\|_F^2/(MN)$
            \STATE Compute $\mathrm{CRB}_{H}$ using \eqref{eq:crb_mse_memoryless}
        \ENDFOR
    \ENDFOR
\ENDFOR
\STATE \textbf{return} averaged $\mathrm{MSE}_{H}$ and $\mathrm{CRB}_{H}$ curves
\end{algorithmic}
\end{algorithm}

\begin{figure}[!b]
\centering
\safeincludegraphics[width=0.95\linewidth]{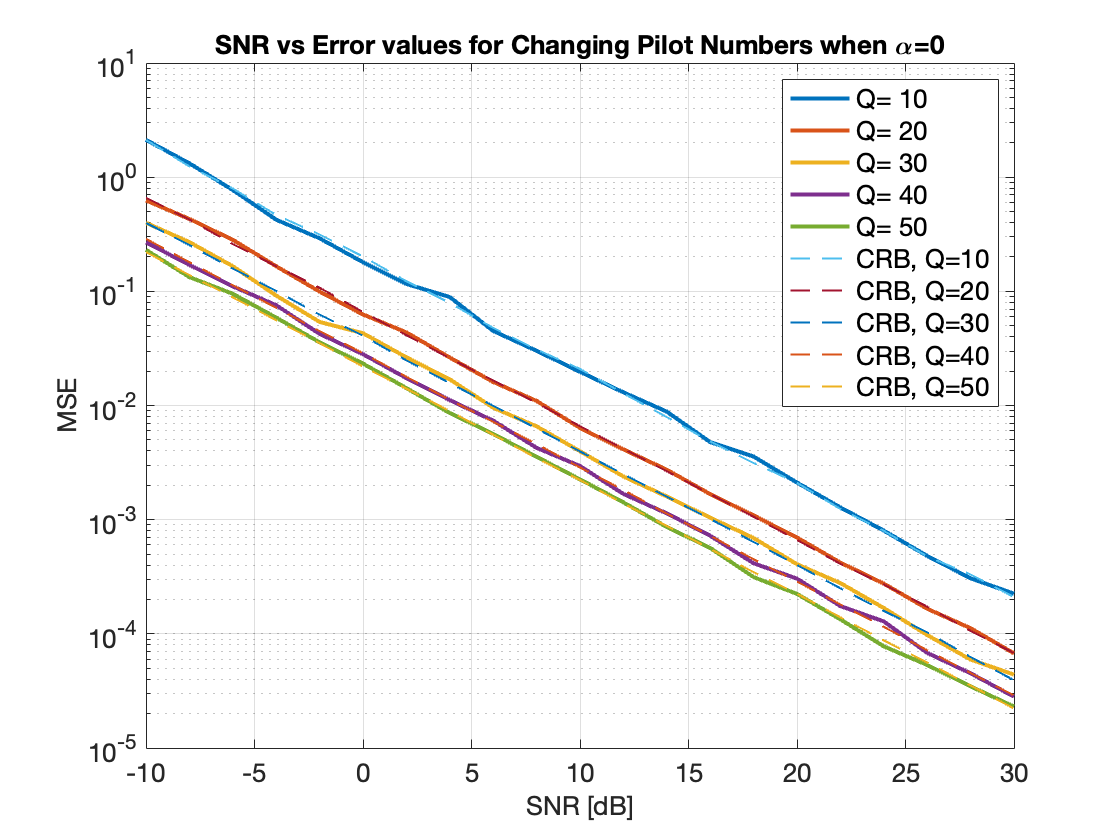}
\caption{Memoryless channel-estimation MSE versus SNR for fixed $\alpha=0$ and different pilot lengths $Q$. The solid curves represent the empirical ML/LS error, and the dashed curves represent the corresponding CRB. Longer pilot blocks improve the conditioning of $\mat X_Q\mat X_Q^T$, lower the CRB, and allow the estimator to approach the theoretical limit over the full SNR range.}
\label{fig:mem_q}
\end{figure}

Fig.~\ref{fig:mem_q} demonstrates the benefit of increasing the number of pilot symbols in the memoryless setting. When $Q$ is small, the normal matrix $\mat X_Q\mat X_Q^T$ is formed from fewer observations, and the resulting estimator has a larger variance. As $Q$ increases, the pilot block provides more independent equations for the sixteen channel coefficients, which reduces the MSE at every SNR. The nearly parallel decrease of the curves with SNR is consistent with the Gaussian linear model, in which reducing the noise variance by a fixed factor reduces the estimation variance by the same factor. The close agreement between the empirical MSE and the CRB confirms that the ML/LS estimator is efficient for this model when the pilot matrix is full rank and well conditioned.

\begin{figure}[!h]
\centering
\safeincludegraphics[width=0.95\linewidth]{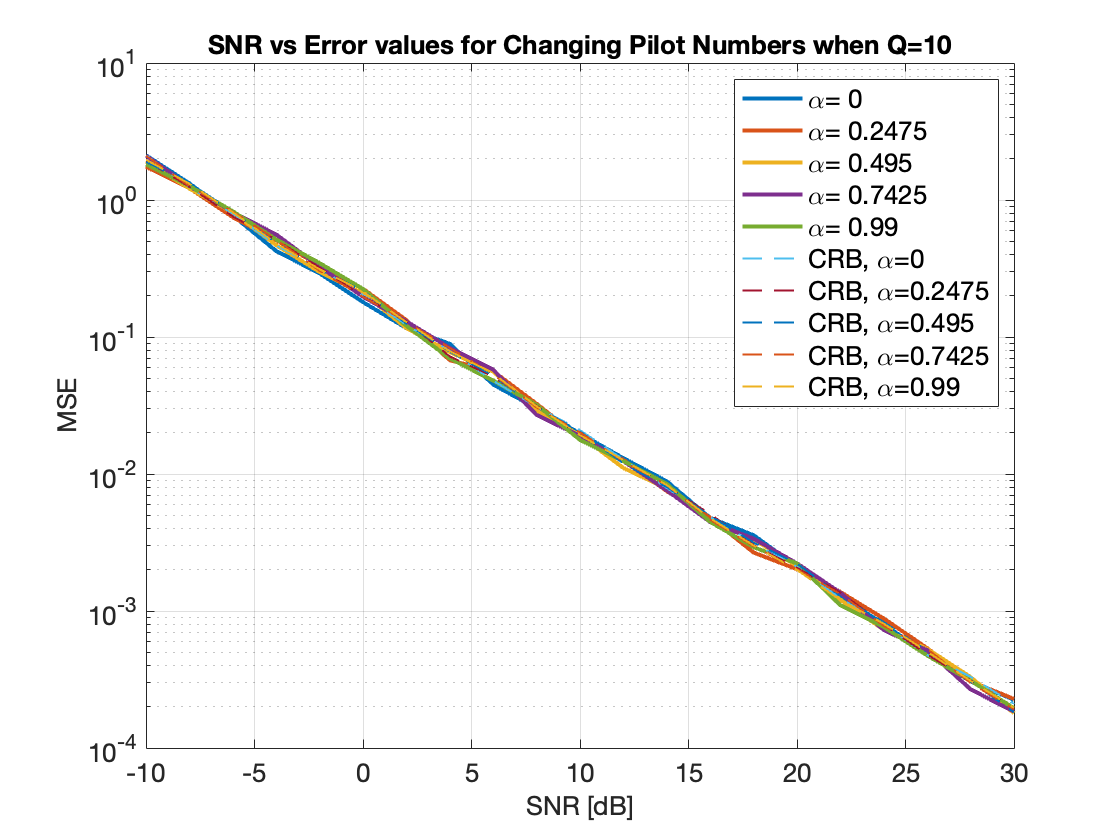}
\caption{Memoryless channel-estimation MSE versus SNR for fixed $Q=10$ and different spatial coupling factors $\alpha$. The curves are nearly overlapping because the estimation variance depends primarily on the pilot matrix and noise covariance, not on the deterministic amplitude pattern of the channel.}
\label{fig:mem_alpha}
\end{figure}

Fig.~\ref{fig:mem_alpha} evaluates the effect of the spatial coupling factor. The results show that varying $\alpha$ does not significantly change the MSE of the memoryless estimator. This behavior is expected from \eqref{eq:fim_memoryless}: for a deterministic unknown channel observed through additive Gaussian noise, the FIM is determined by the pilot matrix and the noise covariance. The actual value of $\mat H$ affects the noiseless received signal, but it does not change the estimator variance under the linear model. This result is useful because it separates channel-identification difficulty from channel-realization magnitude in the memoryless experiment.

\subsection{Four-Tap Channel-Estimation Results}
\label{subsec:fir_results}

Algorithm~\ref{alg:fir_estimation} extends the previous estimator to a four-tap channel. The pilot matrix is expanded into a block convolution matrix so that delayed versions of all transmit streams contribute to the stacked observation vector. This step increases the number of unknown parameters from $MN=16$ to $MNL_h=64$, and it makes the rank of the convolution matrix the dominant condition for successful estimation.

\begin{algorithm}[!t]
\caption{Four-tap MIMO channel estimation and CRB evaluation}
\label{alg:fir_estimation}
\footnotesize
\begin{algorithmic}[1]
\STATE \textbf{Input:} SNR grid $\Gamma$, pilot grid $\mathcal{Q}$, tap length $L_h=4$, parameters $(\alpha,\beta)$, covariance $\mat C_w$
\STATE \textbf{Output:} empirical FIR-channel MSE and CRB
\FOR{each $\gamma\in\Gamma$}
    \FOR{each $Q\in\mathcal{Q}$}
        \STATE Draw pilots $\{\mat x[1],\ldots,\mat x[Q]\}$
        \STATE Construct the block convolution matrix $\mat\Phi_Q$ from delayed pilot vectors
        \FOR{$\ell=0$ to $L_h-1$}
            \STATE Form $[\mat H[\ell]]_{ij}=\alpha^{|i-j|}\beta^{\ell}$
        \ENDFOR
        \STATE Stack all channel taps in $\vect\theta=\vecop([\mat H[0],\ldots,\mat H[L_h-1]])$
        \STATE Form the stacked noise covariance $\mat\Sigma_w=\mat I\otimes\mat C_w/\gamma$
        \STATE Generate $\mat y_Q=\mat\Phi_Q\vect\theta+\mat w_Q$
        \IF{$\mat\Phi_Q^T\mat\Sigma_w^{-1}\mat\Phi_Q$ is nonsingular}
            \STATE Compute $\widehat{\vect\theta}$ using \eqref{eq:fir_gls}
            \STATE Compute $\mathrm{MSE}_{H}=\|\widehat{\vect\theta}-\vect\theta\|_2^2/(MNL_h)$
            \STATE Compute the CRB using \eqref{eq:fir_crb}
        \ELSE
            \STATE Mark the trial as rank-deficient and exclude the CRB curve for that setting
        \ENDIF
    \ENDFOR
\ENDFOR
\STATE \textbf{return} averaged FIR-channel MSE and CRB curves
\end{algorithmic}
\end{algorithm}

\begin{figure}[!t]
\centering
\safeincludegraphics[width=0.95\linewidth]{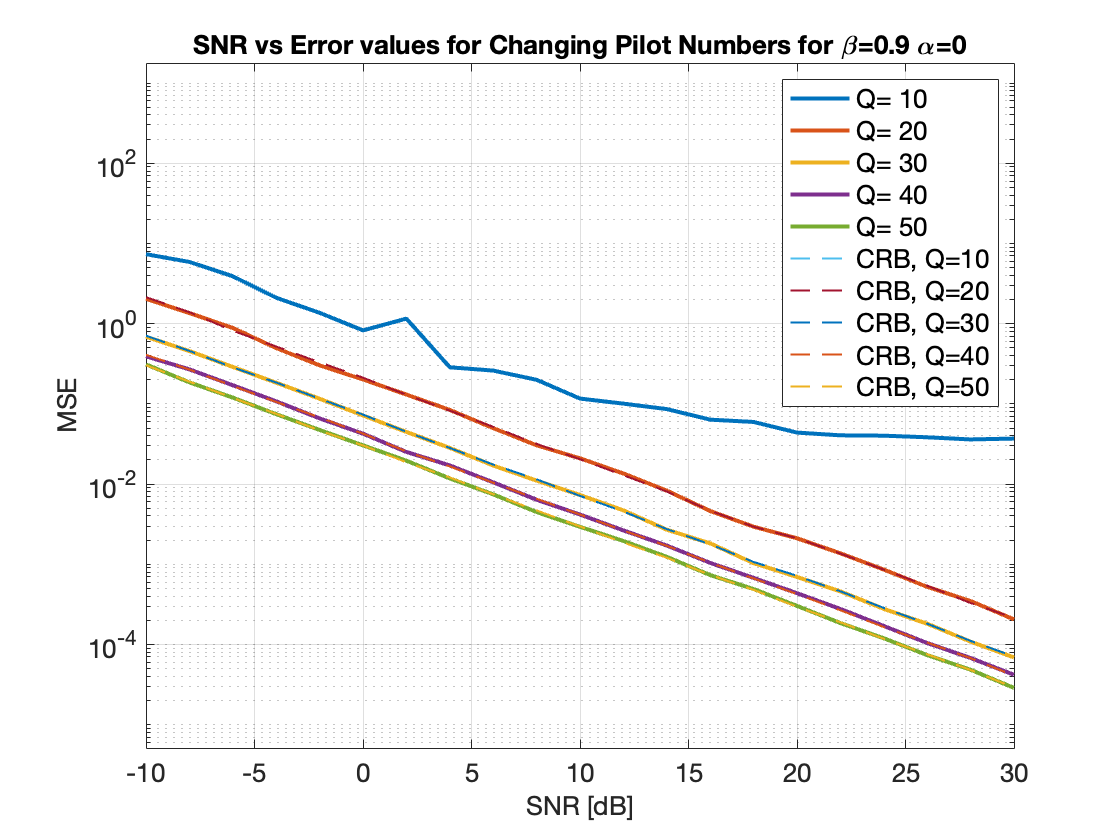}
\caption{Four-tap channel-estimation MSE versus SNR for fixed $\alpha=0$ and $\beta=0.9$. Because each receive branch must identify $NL_h=16$ coefficients, the FIR estimator requires longer pilot sequences than the memoryless estimator. For sufficiently large $Q$, the empirical MSE follows the CRB closely; for short pilots, rank deficiency and poor conditioning increase the error.}
\label{fig:fir_q}
\end{figure}

Fig.~\ref{fig:fir_q} shows the effect of pilot length for the four-tap model. The main difference relative to Fig.~\ref{fig:mem_q} is the substantially larger parameter dimension. Even though the channel remains linear in its unknown coefficients, the convolution matrix must contain enough independent delayed pilot vectors to identify all taps. The shortest pilot configurations produce the largest MSE and may not yield a valid CRB when the normal matrix is singular. As $Q$ increases, the convolution matrix becomes better conditioned, the estimator approaches the CRB, and the error decreases steadily with SNR. This result highlights that channel memory directly increases the training requirement.

\begin{figure}[!t]
\centering
\safeincludegraphics[width=0.95\linewidth]{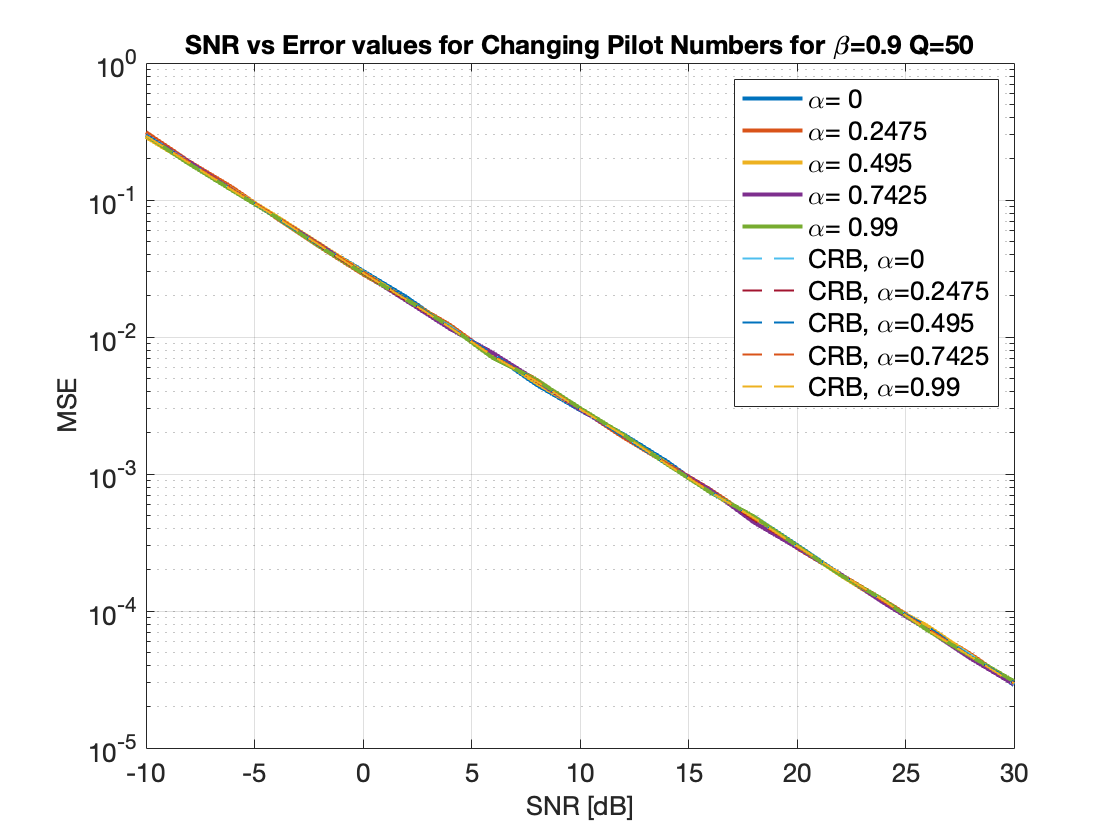}
\caption{Four-tap channel-estimation MSE versus SNR for fixed $Q=50$ and $\beta=0.9$ while varying $\alpha$. The nearly identical curves show that, once the pilot convolution matrix has adequate rank, the deterministic spatial coupling coefficient has limited influence on estimation variance.}
\label{fig:fir_alpha}
\end{figure}

Fig.~\ref{fig:fir_alpha} confirms that the spatial coupling coefficient has little effect on the FIR estimator variance when the pilot block is sufficiently long. The curves remain close to each other for all tested values of $\alpha$, indicating that the dominant factors are SNR, pilot length, and the rank of $\mat\Phi_Q$. The interpretation is similar to the memoryless case, but the FIR setting makes the conditioning requirement more important. The figure also supports the use of CRB comparison as a diagnostic tool: when the MSE follows the bound, the estimator is operating in the information-limited regime rather than being dominated by numerical failure.

\begin{figure}[!b]
\centering
\safeincludegraphics[width=0.95\linewidth]{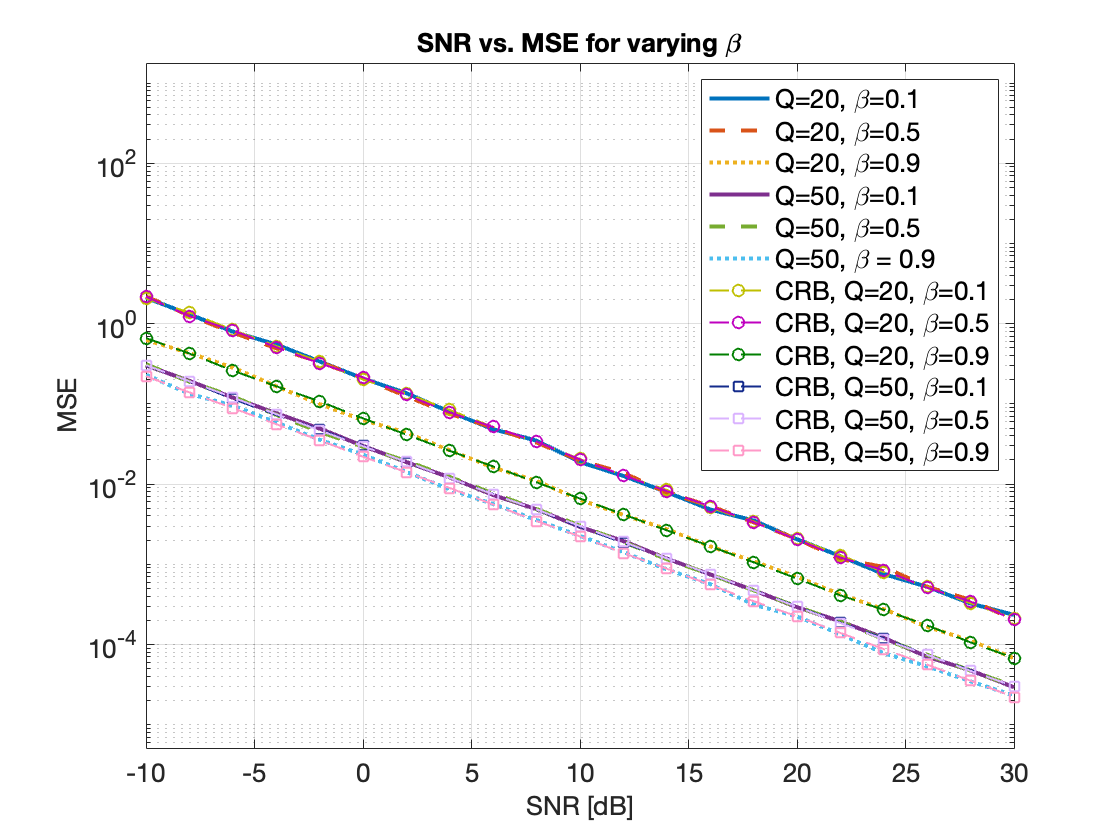}
\caption{Effect of the memory-decay coefficient $\beta$ on four-tap channel-estimation MSE with fixed $\alpha=0$. The tested values represent rapidly decaying, moderately decaying, and slowly decaying impulse responses. In the reported simulation, larger $\beta$ values produce stronger later taps and lead to lower normalized estimation error for the same SNR and pilot length.}
\label{fig:fir_beta}
\end{figure}

Fig.~\ref{fig:fir_beta} isolates the effect of the temporal decay factor. When $\beta$ is small, the later taps have low amplitude, so their estimation is strongly affected by additive noise. When $\beta$ is larger, the impulse response has more significant energy across the four taps, and the normalized coefficient error decreases in the plotted experiments. This behavior is expected because a slower tap decay improves the effective observability of the weaker delayed components, making the channel vector less dominated by only the first few coefficients. The figure also shows that increasing $Q$ improves performance for each value of $\beta$, which confirms that pilot length remains a primary design parameter regardless of the tap-decay profile. In particular, longer pilot sequences provide additional measurements for the estimator, reducing the sensitivity to noise and yielding more reliable recovery of both dominant and weak multipath coefficients. Thus, the results indicate that both the energy distribution among taps and the available pilot resources jointly determine the achievable estimation accuracy.

\subsection{Deconvolution and Pilot-Length Selection}
\label{subsec:deconvolution_results}
\vspace{8pt}
Algorithm~\ref{alg:deconvolution} describes the deconvolution experiment. A frame of length $P=200$ is divided into a pilot portion and a data portion. The pilot portion estimates the channel, and the data portion is recovered using both the ML/zero-forcing inverse in \eqref{eq:zf_deconv} and the MMSE inverse in \eqref{eq:mmse_deconv}. This experiment is designed to show that better channel estimation does not automatically produce the best overall frame design, because increasing $Q$ also reduces the number of available data samples.

\begin{algorithm}[!t]
\caption{Pilot-aided memoryless deconvolution with ML and MMSE inverses}
\label{alg:deconvolution}
\footnotesize
\begin{algorithmic}[1]
\STATE \textbf{Input:} total frame length $P$, pilot grid $\mathcal{Q}$, SNR grid $\Gamma$, channel $\mat H$, covariance $\mat C_w$
\STATE \textbf{Output:} $\mathrm{MSE}_{X,\mathrm{ML}}$, $\mathrm{MSE}_{X,\mathrm{MMSE}}$, and $J(Q)$
\FOR{each $\gamma\in\Gamma$}
    \FOR{each $Q\in\mathcal{Q}$}
        \STATE Draw a frame $\mat X=[\mat X_p,\mat X_d]$ with $Q$ pilot samples and $P-Q$ data samples
        \STATE Generate pilot observations $\mat Y_p=\mat H\mat X_p+\mat W_p$
        \STATE Estimate $\widehat{\mat H}=\mat Y_p\mat X_p^T(\mat X_p\mat X_p^T)^{-1}$
        \STATE Generate data observations $\mat Y_d=\mat H\mat X_d+\mat W_d$
        \STATE Form $\mat A=\widehat{\mat H}^T(\mat C_w/\gamma)^{-1}\widehat{\mat H}$ and $\mat B=\widehat{\mat H}^T(\mat C_w/\gamma)^{-1}\mat Y_d$
        \STATE Recover $\widehat{\mat X}_{\mathrm{ML}}=\mat A^{-1}\mat B$
        \STATE Recover $\widehat{\mat X}_{\mathrm{MMSE}}=(\mat A+\mat I_N)^{-1}\mat B$
        \STATE Compute $\mathrm{MSE}_{X,\mathrm{ML}}$ and $\mathrm{MSE}_{X,\mathrm{MMSE}}$ from \eqref{eq:mse_X}
        \STATE Compute $J(Q)=\mathrm{MSE}_{H}+\mathrm{MSE}_{X,\mathrm{MMSE}}$
    \ENDFOR
\ENDFOR
\STATE \textbf{return} averaged deconvolution-error curves and decision metric
\end{algorithmic}
\end{algorithm}

\begin{figure}[!t]
\centering
\safeincludegraphics[width=0.95\linewidth]{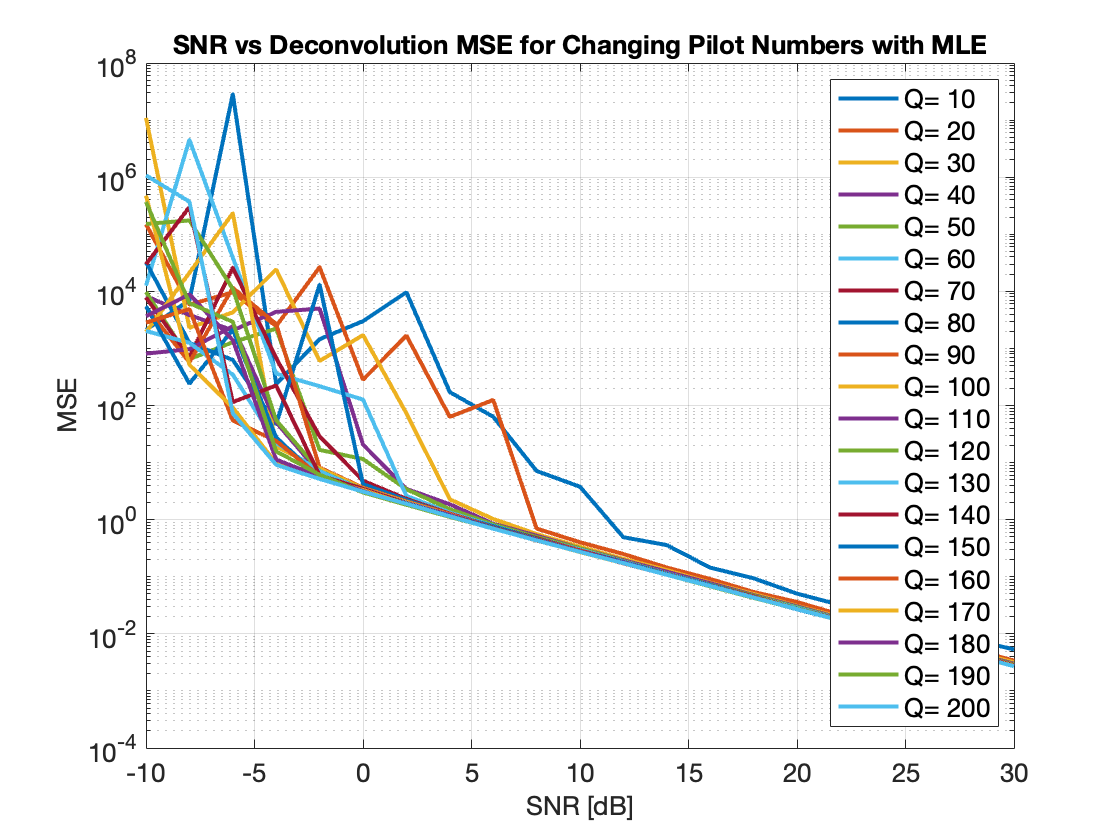}
\caption{Data-symbol reconstruction MSE for ML/zero-forcing deconvolution with different pilot lengths $Q$. The estimator exactly inverts the estimated channel in a weighted LS sense, but it is highly sensitive to channel-estimation error and small singular values. This sensitivity produces large fluctuations and high MSE in the low-SNR and short-pilot regimes.}
\label{fig:deconv_mle}
\end{figure}

Fig.~\ref{fig:deconv_mle} shows the reconstruction performance of the ML/zero-forcing inverse. At high SNR, the channel estimate improves and the reconstruction error decreases. At low SNR, however, the inverse can strongly amplify both thermal noise and channel-estimation error. The irregular behavior for short pilot lengths is a typical consequence of inverting a poorly estimated channel matrix. Increasing $Q$ generally reduces this instability because the estimated channel becomes closer to the true channel, but zero forcing still lacks an explicit mechanism for limiting noise enhancement.

\begin{figure}[!t]
\centering
\safeincludegraphics[width=0.95\linewidth]{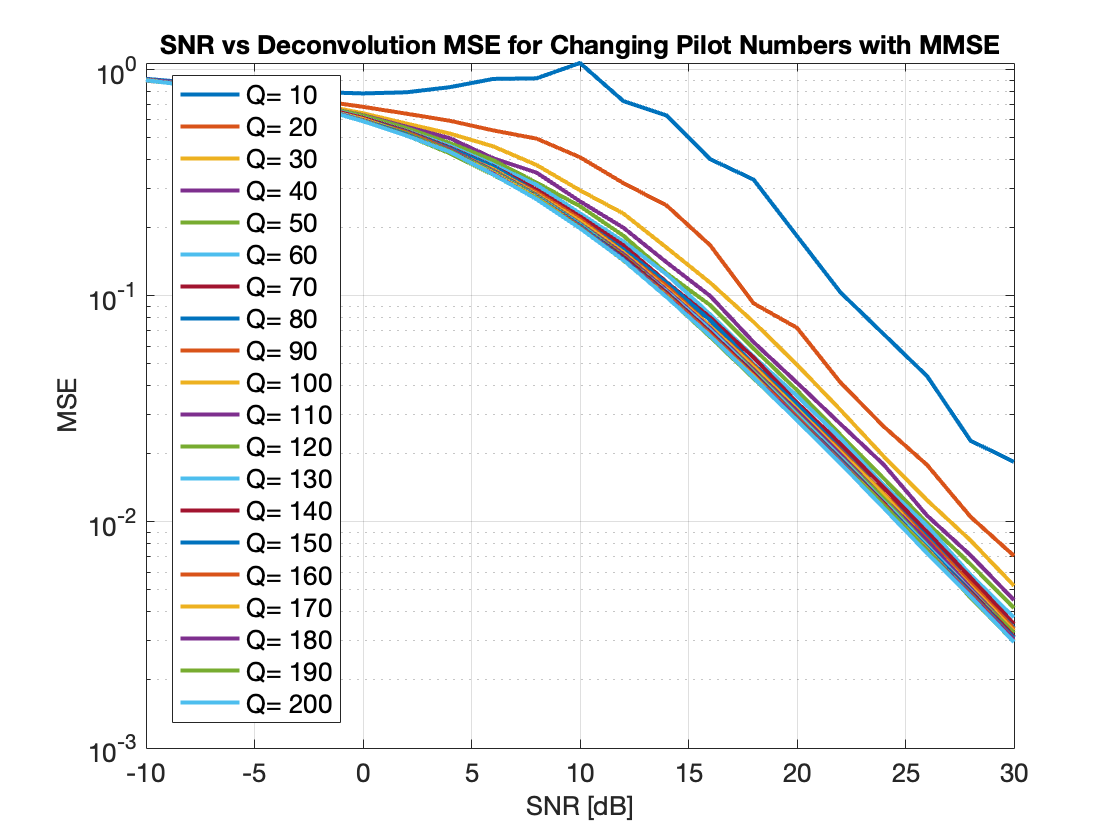}
\caption{Data-symbol reconstruction MSE for MMSE deconvolution with different pilot lengths $Q$. The MMSE estimator uses the same estimated channel as the ML inverse, but the additional identity regularization term in \eqref{eq:mmse_deconv} limits noise amplification. The resulting curves are smoother and more stable, especially for low SNR and limited training.}
\label{fig:deconv_mmse}
\end{figure}

Fig.~\ref{fig:deconv_mmse} reports the MMSE reconstruction error. Compared with the ML curves, the MMSE curves are more regular and maintain lower error in the difficult low-SNR region. This behavior follows directly from the structure of \eqref{eq:mmse_deconv}: the regularization term prevents the inverse from becoming excessively large when $\widehat{\mat H}^{T}\mat C_w^{-1}\widehat{\mat H}$ is ill conditioned. The advantage of MMSE deconvolution becomes less pronounced at high SNR, where the channel estimate is more accurate and the effective noise variance is lower. Nevertheless, the MMSE result provides a more reliable design choice when the available pilot budget is limited.

\begin{figure}[!t]
\centering
\safeincludegraphics[width=0.95\linewidth]{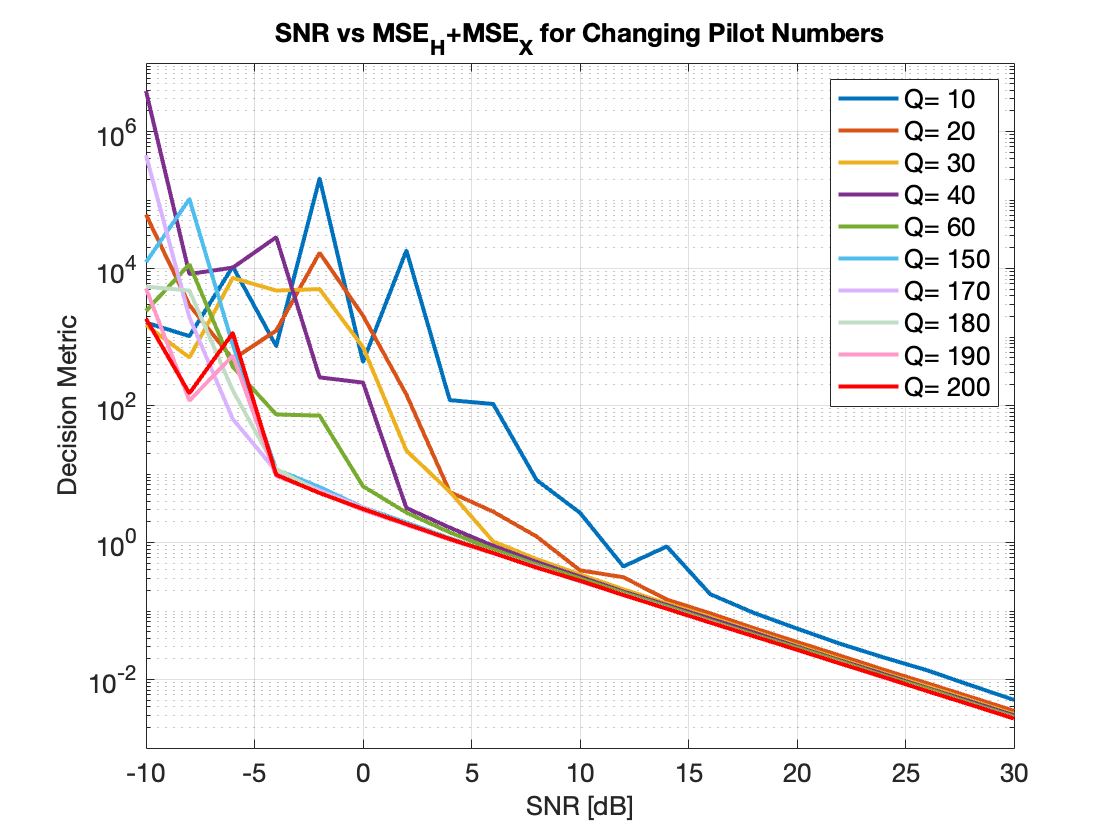}
\caption{Aggregate decision metric $J(Q)=\mathrm{MSE}_{H}+\mathrm{MSE}_{X}$ for pilot-length selection. It combines the channel-estimation error and the data-reconstruction error, so it identifies the pilot length that gives the lowest accuracy loss for each SNR. The result illustrates the tradeoff between spending more samples on training and retaining enough data samples in a fixed-length frame.}
\label{fig:deconv_metric}
\end{figure}

Fig.~\ref{fig:deconv_metric} combines channel and data errors into a single pilot-selection metric. For very small $Q$, the channel estimate is inaccurate, and the reconstruction error is therefore high. Increasing $Q$ improves the channel estimate and reduces the deconvolution error; however, in a fixed-length frame, a larger pilot block leaves fewer samples for data transmission. The metric in \eqref{eq:decision_metric} summarizes this accuracy tradeoff and allows different SNR regimes to be compared on the same plot. In a complete communication-system design, this metric could be combined with spectral-efficiency and coding-rate considerations, but it already gives a useful first-order criterion for selecting the training length.

\section{Conclusion}
\label{sec:conclusion}

This study presented a structured study of pilot-aided MIMO system identification and linear deconvolution under spatially correlated Gaussian noise. The covariance-validation experiments confirmed that the generated noise follows the intended Toeplitz covariance model and that longer sample records are necessary when the noise correlation is strong. The memoryless channel-estimation experiments showed that the ML/LS estimator approaches the CRB when the pilot matrix is full rank and sufficiently well conditioned. The four-tap experiments demonstrated the additional training burden introduced by channel memory; the number of unknown coefficients increases by a factor of four, and short pilot sequences can lead to singular or poorly conditioned estimation problems.

The four-tap experiments demonstrated the additional training burden introduced by channel memory; the number of unknown coefficients increases by a factor of four, and short pilot sequences can lead to singular or poorly conditioned estimation problems. Compared with the memoryless channel, the multipath channel requires substantially more pilot resources to ensure reliable parameter estimation and stable matrix inversion. The results showed that insufficient pilot lengths produce larger coefficient-estimation errors and increase the variability of the estimates across independent trials. As the pilot length grows, however, the estimation accuracy improves significantly and the performance gap relative to the memoryless case narrows. These observations illustrate a fundamental trade-off in practical MIMO systems: richer channel models provide a more realistic representation of multipath propagation but require additional training overhead and computational complexity. The experiments therefore emphasize the importance of selecting pilot lengths that scale appropriately with the channel memory in order to maintain reliable system identification and avoid numerical instability.

The simulation results also highlight several practical trade-offs encountered in real communication systems. Increasing the pilot length generally improves channel-estimation accuracy and reduces deconvolution error, but it also consumes transmission resources that could otherwise be allocated to payload data. Similarly, channels with longer memory require more parameters to be estimated, increasing both the computational burden and the training overhead. These observations demonstrate that reliable system identification requires a careful balance between estimation accuracy, spectral efficiency, and implementation complexity.

Another important outcome of this study is the close relationship between channel-estimation performance and subsequent signal recovery. Even moderate estimation errors can significantly affect the quality of deconvolution, particularly when the channel matrix is poorly conditioned. Consequently, the design of pilot sequences and estimation algorithms should not be viewed independently from the final detection or reconstruction stage. Instead, the entire estimation-and-equalization chain should be optimized as a unified system to achieve robust performance under realistic noise and channel conditions.

Future work could extend the analysis to frequency-selective fading channels with larger delay spreads, time-varying MIMO channels, and non-Gaussian interference environments. Additional investigations could also compare ML/LS estimators with Bayesian, sparse-recovery, or adaptive estimation techniques that exploit prior channel structure. Furthermore, the incorporation of advanced regularization methods and iterative deconvolution algorithms may provide additional robustness in challenging low-SNR scenarios. Such extensions would further enhance the understanding of MIMO system identification and contribute to the development of reliable channel-estimation and signal-recovery techniques for next-generation wireless communication systems.

\end{document}